\documentclass[aps,twocolumn,showpacs,showkeys,preprintnumbers,amsmath,amssymb]{revtex4}

\usepackage{epsfig}

\usepackage{dcolumn}
\usepackage{amsmath}

\usepackage{graphicx}
\usepackage{rotating}
\usepackage{multirow}

\def\be{\begin{equation}}
\def\ee{\end{equation}}
\def\lsim{\raise0.3ex\hbox{$<$\kern-0.75em\raise-1.1ex\hbox{$\sim$}}}
\def\gsim{\raise0.3ex\hbox{$>$\kern-0.75em\raise-1.1ex\hbox{$\sim$}}}

\begin{document}

\newlength{\figurewidth}
\ifdim\columnwidth<10.5cm
  \setlength{\figurewidth}{0.95\columnwidth}
\else
  \setlength{\figurewidth}{10cm}
\fi
\setlength{\parskip}{0pt}
\setlength{\tabcolsep}{6pt}
\setlength{\arraycolsep}{2pt}

\title{Do extremists impose the structure of social networks?}

\medskip
\author{Philippe Blanchard}

\vskip0.3cm

\author{Santo Fortunato}

\vskip0.5cm

\affiliation{Fakult\"at f\"ur Physik $\&$ BiBoS, Universit\"at Bielefeld, 
D-33501 Bielefeld, Germany}

\vskip0.3cm

\author{Tyll Kr\"uger}

\vskip0.5cm

\affiliation{Fakult\"at f\"ur Mathematik $\&$ BiBoS, Universit\"at Bielefeld, 
D-33501 Bielefeld, Germany}

\begin{abstract}
\noindent

The structure and the properties of complex networks essentially depend
on the way how nodes get connected to each other. 
We assume here that each node has a feature which attracts the others.
We model the situation by assigning 
two numbers to each node, $\omega$ and $\alpha$, where 
$\omega$ indicates some property 
of the node and $\alpha$ the affinity towards that property. 
A node $A$ is more likely to establish a connection with 
a node $B$ if $B$ has a high value of $\omega$ and $A$ has a high 
value of $\alpha$.
Simple computer simulations show
that networks built according to this principle have a degree distribution
with a power law tail, whose exponent is determined
only by the nodes with
the largest value of the affinity $\alpha$ (the "extremists").
This means that the extremists lead the formation process 
of the network and manage to shape the final topology of the system.
The latter phenomenon may have implications 
in the study of social networks and in epidemiology.

\end{abstract}

\pacs{89.75.Hc, 05.10.-a}

\keywords{Complex networks, degree distribution, extremists.}

\maketitle

\vskip0.7cm

The study of complex networks is currently one of the hottest fields 
of modern physics \cite{linked, bara,newman}. 
A network (or graph) is a set of items, called {\it vertices} or {\it nodes}, 
with connections between them, called {\it edges}. Nodes linked 
by an edge are neighbours and the number of neighbours 
of a node is called {\it degree}. Complex weblike 
structures describe a wide variety of systems of high 
technological and intellectual importance.
Examples are the Internet, the World Wide Web (WWW), social networks of 
acquaintance or other connections between individuals, neural networks,
food webs, citation networks and many others. 

One of the crucial questions concerns the formation of these structures.
Complex networks are in general systems in evolution,
with new nodes/edges that get formed and old ones that get 
removed or destroyed. 
The currently accepted mechanism finds its roots 
in an old idea of Price \cite{price}, based on the so-called 
preferential 
attachment, which means that a newly formed node $A$ builds an edge with a
preexisting node with a probability that is proportional to the degree of the 
latter node. Networks constructed in this way \cite{aba, doro1,redner}
have a degree distribution with a power law
tail, as observed in real networks.
This simple rule, however, makes implicitely the strong assumption that each
node is at any time informed about the degree of all other nodes, which is 
certainly not true, especially for gigantic systems which 
contain many millions of nodes, like the WWW. We rather believe that 
the key behind the building of a connection between a pair of nodes 
lies essentially in the mutual interaction of the two nodes, (almost) 
independently of the rest of the system: two persons usually become friends because 
they like each other. 

In this letter we have social networks in mind, but nevertheless 
we will speak generally about networks, as we believe that our model
has a more general validity.
The mechanism we propose is that 
any node has some {\it property} (beauty, richness, power, etc.)
by which the others are {\it attracted}.
We indicate the property with a positive number $\omega$, the attractiveness by another
positive number $\alpha$. We assume that high values of $\omega$ correspond to a high
degree of the property (the most beautiful people, for instance).
Both $\omega$ and $\alpha$ are attributes of single nodes/individuals, so they 
take in general different values for different nodes. What we need is a
knowledge of the distribution of $\omega$ and $\alpha$ in the network.
For the property $\omega$, distributions which vanish for high values of
$\omega$, like exponentials or power laws, 
are realistic. As far as the affinity $\alpha$ 
is concerned, it is less clear
which distributions can be considered plausible, therefore we tested several functions.
We remark that the idea that the nodes
have individual appeal already exists in the literature
on complex networks. Bianconi and Barab\'asi \cite{bianconi} assigned a parameter
$\eta$ called "fitness" to each node of the network
and the linking probability 
becomes proportional to the product of the degree with the fitness of the target node.
In the same framework, 
Erg{\"u}n and Rodgers \cite{ergun} proposed a different ansatz for the 
linking probability, which in their case is proportional to the sum of the fitness and
the degree of the target node. Both models however are based on preferential
attachment. Caldarelli et al. \cite{calda} proposed instead a variation of the fitness theme
which eliminates preferential attachment, so that the formation principle of networks 
lies in the attraction which nodes exert on each other by virtue of their
individual quality/importance, which is actually in the spirit of our work.
So, in \cite{calda}, the linking probability is simply
a function of the fitness values of the pair of nodes, and several
possible choices for this function are introduced and discussed.

Our expression for the probability $p_{AB}$ of a node $A$ to build
an edge with a node $B$ is also a function of the individual attributes of the
nodes, $\omega$ and $\alpha$. We adopt the ansatz

\begin{equation}
p_{AB}=\frac{c_B}{[\phi(\omega_B)]^{\alpha_A}},
\label{eq1}
\end{equation}

where $c_B$ is a normalization constant and 
$\phi(\omega)$ the distribution function of the property
$\omega$, whereas we will indicate with $\psi(\alpha)$ the distribution
function of the affinity $\alpha$. 
What (\ref{eq1}) says is that the pairing probability 
is inversely proportional to the relative frequency of the property 
$\omega$ in the network. Thinking again of a social system,
the idea is that there is a
general tendency to be more attracted by those subjects who are
characterized by high values of $\omega$. In a network of sexual relationships,
for instance, the best looking people usually have the greatest chances to be 
chosen as sexual partners.
We believe that
the choices of the people are not influenced by
the absolute importance of $\omega$, which 
is a vague and abstract concept, but rather 
by the perception of the importance of the property $\omega$ within the society,
which is related to its distribution.
This is why we 
associated the pairing probability to the relative frequency $\phi(\omega)$ 
and not directly to $\omega$, at variance with \cite{calda}.
Accordingly, the larger $\omega$, the lower the occurrence $\phi(\omega)$ of that 
degree of the property in the network, and 
the edge-building probability gets higher.
On the other hand, for a given node $B$, characterized by its property
$\omega_B$, the other nodes will feel an attraction towards $B$ which varies 
from a subject to another. This modulation of the individual 
attraction is expressed by the exponent $\alpha_A$ in (\ref{eq1}).  
The probability $p_{AB}$ 
increases with $\alpha_A$, justifying the denomination 
of "affinity" we have given to the parameter $\alpha$. 
The parameter $\alpha$ must be taken in the range $[0,1]$ for normalization
purposes \cite{nota1}.

The expression (\ref{eq1}) might look {\it ad hoc}, because of the 
power law dependence on $\alpha$. This is not
the case, due to the freedom we have in the choice of the $\alpha$'s: 
if we have an arbitrary probability $p_{AB}$ for 
the node $A$ to be linked to $B$, in most 
cases one is able to find a number $\alpha_A$ 
such that $p_{AB}=c_B/[\phi(\omega_B)]^{\alpha_A}$, so to reproduce the wished probability. 

Our simple model is a generalization of the so-called
"Cameo principle" which has recently been introduced by two of the authors 
\cite{cameo}. There, the
affinity $\alpha$ was the same for all nodes and there was consequently no
correlation between pairs of nodes.
In this case it was rigorously proven that the network has indeed a degree
distribution with a power law tail and that the exponent $\gamma$ 
is a simple function of $\alpha$, more precisely

\begin{itemize}
\item{if $\phi(\omega)$ decreases as a power law with exponent 
$\beta$ when $\omega\rightarrow\infty$, $\gamma=1+1/\alpha-1/{\alpha\beta}$;}
\item{if $\phi(\omega)$ vanishes faster than any power law when
    $\omega\rightarrow\infty$, $\gamma=1+1/{\alpha}$.}
\end{itemize}
 
\begin{figure}[htb]
  \begin{center}
    \epsfig{file=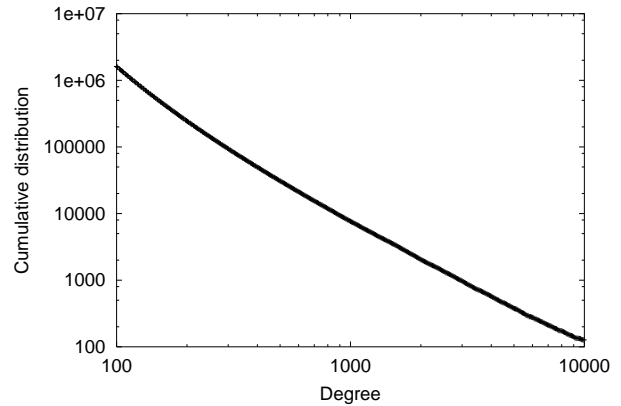,width=9cm}
    \caption{\label{fig1}{Cumulative degree distribution for a network where
$\phi(w)=e^{-\omega}$ and $\psi(\alpha)$ is constant in $[0,0.7]$. In the double
logarithmic scale of the plot a power law tail would appear as a straight line,
as in the figure.}}
  \end{center}
\end{figure}

For our generalization we studied the problem numerically,
by means of Monte Carlo simulations, but an analytical proof of the results
we show here is in progress \cite{rancameo}. 
In order to build the network 
we pick up a node $A$ and build $m$ edges with the other nodes 
of the network, with probability given by (\ref{eq1}). 
The procedure is then repeated for all other nodes of the network.
We remark that our construction process is static, i. e. all nodes of the
network are there from the beginning of the process and neither nodes are added
nor destroyed. However, the principle can as well be implemented 
in a dynamical way, with new nodes which are progressively added to the network
\cite{cameo}. 

We fixed the outdegree 
$m$ to the same value for all nodes, as it is done in 
the famous model of Barab\'asi and Albert \cite{aba} (we set $m=100$).
The number $N$ of nodes was mostly $100000$.

We have always used a simple exponential for $\phi(\omega)$.
Fig. \ref{fig1} shows the cumulative degree distribution of the network
constructed with a uniform affinity distribution $\psi(\alpha)=const.$,
for $\alpha$ in the range $[0,0.7]$. The cumulative distribution is the 
integral of the normal distribution. So, for a value $k$ of the degree 
we counted how many nodes have degree larger than $k$. The summation 
reduces considerably fluctuations and the analysis gets easier.
If the degree distribution is a power law with exponent $\gamma$, 
its integral will be again a power law but with exponent $\gamma-1$.

In Fig. \ref{fig1} we see that indeed the cumulative distribution ends 
as a straight line in a double logarithmic plot, so it has a power law tail.
We performed many trials, by 
varying the range of the uniform distribution $\psi(\alpha)$, and by
using other kinds of distribution functions for $\alpha$, like 
gaussians, exponentials and power laws. We found that
the result holds in all cases we considered. 

\begin{figure}[htb]
  \begin{center}
    \epsfig{file=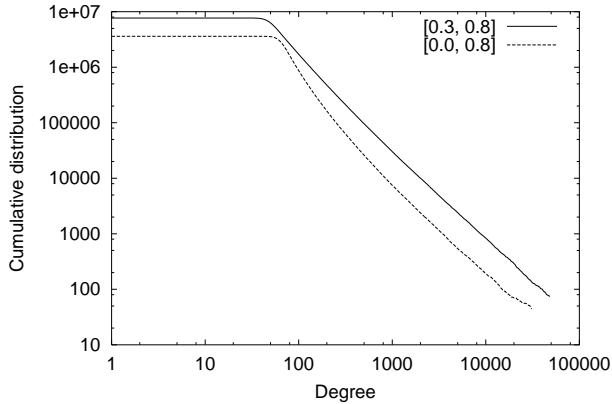,width=9cm}
    \caption{\label{fig2}{Cumulative degree distribution for two networks
characterized by uniform distributions $\psi(\alpha)$. The upper limit of the
$\alpha$-range is the same in both cases. The power law tails
have the same slope.}}
  \end{center}
\end{figure}

Another striking feature of our findings is shown in
Fig. \ref{fig2}. Here we plot the cumulative degree distributions for two networks,
where $\psi(\alpha)$ is uniform and we 
chose the affinity ranges such that 
they share the same upper limit $\alpha_{max}$ ($[0,0.8]$ 
and $[0.3,0.8]$, respectively, so $\alpha_{max}=0.8$).
We see that the tails of the 
two curves have the same slope, which suggests that $\gamma$
only depends on $\alpha_{max}$.
We repeated this experiment several times, for different ranges and 
taking as well exponential and gaussian distributions for $\alpha$.
Within errors, we confirmed this remarkable result.

\begin{figure}[htb]
  \begin{center}
    \epsfig{file=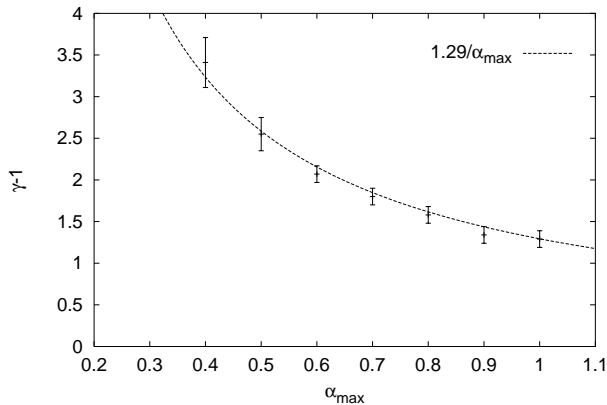,width=9cm}
    \caption{\label{fig3}{Dependence of the exponent $\gamma-1$
of the cumulative degree distribution
on the upper limit $\alpha_{max}$ of the 
affinity range. The data points can be fitted by the simple 
ansatz $a/{\alpha_{max}}$.}}
  \end{center}
\end{figure}

Fig. \ref{fig3} shows how the exponent $\gamma-1$ 
of the cumulative degree distribution varies with $\alpha_{max}$.
The pattern of the data points follows an hyperbola
$a/\alpha_{max}$, with a coefficient $a=1.29$; this is very
close to what one gets for the original Cameo principle \cite{cameo}, 
where $\gamma-1=1/\alpha$. It is likely that 
in the limit of infinite nodes the coefficient would indeed converge 
to one. Since $\alpha_{max}$ can be chosen arbitrarily
close to zero, from the ansatz $a/\alpha_{max}$ we deduce that,
within our model, we are able to build networks with 
any value of $\gamma$ greater than (about) $2$. This is fine, as
for the great majority of complex networks $\gamma{\geq}2$ as well.

So, we have discovered that the nodes with the highest affinity $\alpha$,
that we call "extremists" for obvious reasons,
are responsible for the exponent $\gamma$ of the power law tail of the
degree distribution of the network. This is valid independently of the 
distribution $\psi(\alpha)$, so it works even in the case where 
the extremists are just a very small part of the population \cite{nota2}.
If we consider terrorism networks, for example, the
leaders of the group (those with highest charisma/$\omega$)
are the hubs of the network, i.e. the most connected
individuals, but their relative importance
is determined by the
most fanatic followers (those with largest $\alpha$).
We have then shown that there is a sort of time-dependent hierarchy among the
nodes: the extremists lead
the formation process, the hubs dominate the structure once the network is built.

We give here a hint to the analytical proof of this result.
Let us consider the simple case of a discrete distribution
$\Psi(\alpha)$ of the form

\begin{equation}
\Psi(\alpha)=\sum_{i=1}^{m}{\lambda_i}\delta(\alpha-{\alpha_i}),
\label{eq2}
\end{equation} 

with $\lambda_i>0$ and $\sum_{i=1}^{m}\lambda_i=1$. 
So the fraction of nodes $x$
with $\alpha(x)=\alpha_i$ is $\lambda_i>0$. 
The validity of the result lies in the fact that 
the global degree distribution is given by a superposition 
of the degree distributions associated to nodes with the same $\alpha_i$.
Since each of those distribution has a power law tail \cite{cameo},
the overlap is dominated by the term having the fattest tail, i. e. the
smallest exponent $\gamma$, which corresponds to the maximum $\alpha_{max}$ of the 
$\alpha$'s, due to the relation $\gamma=1+1/\alpha$. 
In this way, any function can be considered as the limit of a sum like (\ref{eq2}),
when the number of terms goes to infinity; for more details see \cite{cameo}
and \cite{rancameo}.

We know that the exponent $\gamma$ is a crucial 
feature of complex networks in many respects. For epidemic spreading,
for example, there is no non-zero epidemic threshold \cite{vespi} so long as 
$\gamma{\leq}3$, which would have catastrophic consequences. If the network is
in evolution, to control the extremists would mean to be able to exert an influence
on the future topology of the network, which can be crucial in many circumstances. 


From a practical point of view, it is not obvious how to model
things like attractiveness (or fitness), 
which usually are out of the domain of quantitative
scientific investigations. 
However, our result
on the leading role of the extremists is quite robust, as it does not depend 
on the specific function $\psi(\alpha)$ that one decides to adopt.
The attempt to mathematically modelize apparently abstract features of
social systems (here the "attractiveness") is not isolated \cite{nota3}.
The last few years witnessed a big effort to describe society as a physical
system \cite{weidlich, ball,st0}, with people playing the role of atoms 
or classical spins undergoing 
elementary interactions. There are meanwhile several models to explain
how hierarchies \cite{bonabeau} and consensus \cite{stauff}
may originate in a society and in these models abstract items like
opinion, confidence, etc. are associated to 
well-defined mathematical variables. Although one must always be careful not to
demand too much from such models, 
the first results of this line of research
are encouraging; with the consensus model of Sznajd \cite{sznajd}
one could reproduce 
the final distribution of votes among 
candidates in Brazilian and Indian elections \cite{st,gonz}.

In conclusion, we have introduced a simple criterion for the nodes
of a complex network to choose each other as terminals of mutual 
connections: each node has a property $\omega$ which attracts the other nodes to
an extent which depends on another individual parameter $\alpha$. 
Networks built in this way are always characterized by
a degree distribution with a power law tail. The exponent of the 
power law is determined uniquely by those nodes which are most sensible to the 
property $\omega$. Acting on such nodes 
could be an effective way to control the 
structure of evolving networks. 

S. F. gratefully acknowledges the financial support of
the TMR network ERBFMRX-CT-970122 and the DFG Forschergruppe FOR
339/1-2; T. K. acknowledges the financial support of the DFG-Forschergruppe FOR 399/2.


\begin{thebibliography}{99}
\bibitem{nota1} In fact, the total probability of a node $A$ to get joined to any of
  the others must be $1$, and the integral of (\ref{eq1}) over all the network
  is convergent only if $\alpha_A<1$.
\bibitem{nota2} It is understood that we refer to the limit of infinitely many
  nodes.
\bibitem{nota3} We emphasize that we are not proposing any concrete model for attractiveness.
 Our viewpoint is rather: "if one would have a model, then ...".
\end{thebibliography}
\end{document}